\documentclass[conference]{IEEEtran}
\IEEEoverridecommandlockouts

\usepackage{cite}
\usepackage{amsmath,amssymb,amsfonts}
\usepackage{algorithmic}
\usepackage{graphicx}
\usepackage{textcomp}
\usepackage{xcolor}

\usepackage{booktabs}
\usepackage{multirow}
\usepackage{array}
\usepackage[hidelinks]{hyperref}
\usepackage{cleveref}
\usepackage{amssymb}
\usepackage{makecell}

\usepackage{siunitx}
\usepackage{fancyhdr}

\usepackage{tikz}

\def\BibTeX{{\rm B\kern-.05em{\sc i\kern-.025em b}\kern-.08em
    T\kern-.1667em\lower.7ex\hbox{E}\kern-.125emX}}
\renewcommand{\baselinestretch}{0.97}
\Crefname{figure}{fig.}{figures}
\Crefname{figure}{Fig.}{Figures}
\crefname{equation}{Eq.}{Eqs.}
\Crefname{equation}{Eq.}{Eqs.}
\def\BibTeX{{\rm B\kern-.05em{\sc i\kern-.025em b}\kern-.08em
    T\kern-.1667em\lower.7ex\hbox{E}\kern-.125emX}}
\begin{document}

\title{ZTA-Q: an Open-source RISC-V Platform for Accurate Quantized CNN Inference

\thanks{This work was supported by Taighde Éireann – Research Ireland (22/US/3848), the National Science Foundation (2315851, 2106634), and the Department for the Economy, Northern Ireland (USI-226).}
}

\author{
\IEEEauthorblockN{
Yike Li\IEEEauthorrefmark{1},
Ajay Kumar M\IEEEauthorrefmark{1},
Vishnu PS\IEEEauthorrefmark{1},
Dimitrios S. Nikolopoulos\IEEEauthorrefmark{3}, \\
Bo Ji\IEEEauthorrefmark{3},
Hans Vandierendonck\IEEEauthorrefmark{4},
Deepu John\IEEEauthorrefmark{1}
}

\IEEEauthorblockA{\IEEEauthorrefmark{1}University College Dublin, Ireland\hspace{1em}%
\IEEEauthorrefmark{3}Virginia Tech, USA\hspace{1em}%
\IEEEauthorrefmark{4}Queen's University Belfast, UK}
}





\maketitle

\begin{abstract}
Low-precision inference is widely adopted in edge AI to reduce computational cost and memory footprint. However, existing open-source accelerator platforms provide limited end-to-end support for CNNs following the standard TensorFlow Lite integer inference scheme. This paper presents ZTA-Q, an open-source RISC-V-based platform that enables accurate deployment of TensorFlow Lite INT8 models. In addition to extending operator support, ZTA-Q provides a configurable post-processing datapath for studying how circuit-level approximations, including reduced multiplier precision, shared shift scaling, and simplified rounding, affect model accuracy. The proposed system is implemented on a Digilent Arty A7-100T FPGA and operates at 83.3 MHz. Evaluations on representative CNN models show that with LUT, register, and DSP overheads of 26.3\%, 12.6\%, and 150\%, respectively, ZTA-Q limits the degradation in both top-1 and top-5 accuracy to within 0.25 percentage points.
\end{abstract}

\begin{IEEEkeywords}
Deep Learning, Hardware accelerators, RISC-V, Quantization, FPGA
\end{IEEEkeywords}
     
\section{Introduction}
AI applications at the edge, such as autonomous vehicles and surveillance cameras, must operate under strict constraints on area, power, and memory. To accelerate computation-intensive operators, recent systems increasingly combine RISC-V processors with dedicated neural processing units (NPUs), providing both general-purpose programmability and efficient tensor computation\cite{kalapothas2023survey}. For open-source RISC-V SoCs, end-to-end model deployment typically involves three stages: a front-end model parser, a domain-specific middle-end compiler, and a target accelerator backend. \Cref{open-source platform} summarizes the deep learning frameworks, including PyTorch(torch), ONNX, TensorFlow(TF), Keras and Caffe, targeted by recent platforms. Several works \cite{genc2021gemmini,yi2025opengemm,kumar2025marvel} rely on third-party tools, such as ONNX Runtime\cite{onnxruntime}, MLIR\cite{lattner2021mlir}, or TVM \cite{chen2018tvm}, to import models from mainstream deep learning frameworks. Although this approach reduces development effort compared to an in-house parser, their multi-layer abstractions can make low-level hardware specification less explicit, which may complicate architectural exploration and extension.


Another limitation lies in the quantization flow, which converts wide accumulator values back to low-bitwidth activations after each layer. In \cite{kumar2025marvel}, TVM-generated C code is executed entirely on the RISC-V cores, without a dedicated hardware implementation for quantization. Other platforms \cite{genc2021gemmini,houshmand2022diana,yi2025opengemm,Marsellus} provide only partial support, implementing either per-tensor or per-channel quantization. These two schemes differ in the granularity at which weight scales are assigned to the whole weight tensor or independently to each channel. The work in \cite{kumar2025bare} integrates NVDLA \cite{nvdla_github}, which provides both per-tensor and per-channel quantization schemes. However, only LeNet-5 and ResNet-18 are evaluated, partly due to the limited support for fully quantized model compilation in the NVDLA software stack.

\begin{table}[t!]
    \centering
    \caption{Deep-learning Framework and quantization support on Open-source RISC-V platforms}
    \label{open-source platform}
    \begin{tabular}{l|c | c |c}
        \hline
        \textbf{Platform} & \textbf{DL Frame} & \textbf{In-house Parser} & \textbf{Quantization} \\
        \hline
        Gemmini \cite{genc2021gemmini}     & ONNX & X   & tensor \\ 
        DIANA   \cite{houshmand2022diana}    & /  & \checkmark     & tensor \\     
        OpenGeMM \cite{yi2025opengemm}   & torch,ONNX & X        & tensor \\
        MARVEL   \cite{kumar2025marvel}   & Keras,torch,TF   & X       & / \\
        \cite{kumar2025bare} & Caffe & \checkmark & tensor/channel \\
        Marsellus \cite{Marsellus}  & torch & \checkmark         & channel \\   
        \hline
    \end{tabular}
\end{table}


\begin{figure*}[!t]
    \centering
    \includegraphics[width=\textwidth]{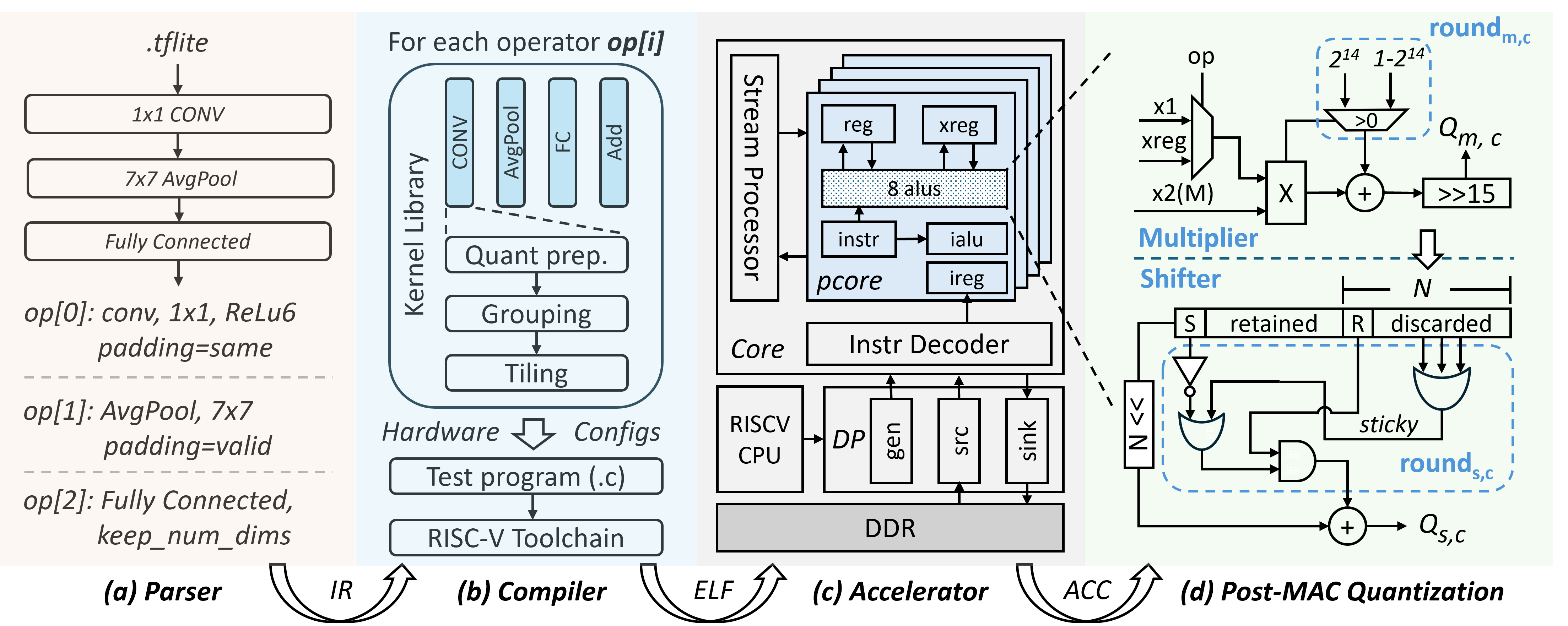}
    \caption{Overview of the proposed ZTA-Q quantized CNN inference flow:
(a) model parsing and IR generation, 
(b) operator compilation through quantization preparation, grouping and tiling, 
(c) execution on the SIMD accelerator, and 
(d) the quantization datapath with multiplier rounding and shift rounding.
    }
    \label{ztachip_arch}
\end{figure*}

To address these two limitations, we propose ZTA-Q, an open-source RISC-V-based CNN inference platform built upon ztachip \cite{ztachip_github}. Through software–hardware co-design, ZTA-Q supports standard TensorFlow Lite integer models and introduces a post-MAC quantization datapath with two-stage rounding, extending the original per-tensor implementation to both per-tensor and per-channel quantization. The main contributions of this work are summarized as follows:

1) We redesign the quantization mapping in the ztachip parser and compiler to provide in-house support for both per-tensor and per-channel quantization. We further introduce a post-MAC datapath with two-stage rounding to implement both schemes directly within the acceleration engine.

2) We provide a per-layer ablation study on how circuit-level approximations in the quantization flow, including reduced scale precision, shared shift scaling, and simplified rounding, affect inference accuracy and hardware cost.

We evaluate ZTA-Q on representative quantized CNN models, including LeNet-5, SqueezeNet1.1, MobileNetV1/V2, and ResNet-18/50. The proposed system is implemented on a Digilent Arty A7-100T board at 83.3 MHz. With our extended quantization scheme support, ZTA-Q achieves accuracy comparable to TensorFlow Lite, with less than 0.25 percentage points of accuracy degradation.

\section{Platform Architecture}
In this section, we first introduce the architecture of the baseline platform ztachip. Then we present how we map per-channel quantization to the ztachip parser and compiler. Finally, we describe our post-MAC quantization flow with rounding in detail.

\subsection{Baseline Platform: ztachip}
Ztachip is an open-source RISC-V AI acceleration platform for edge inference, with an in-house parser and compiler targeting quantized TensorFlow Lite models, as shown in \Cref{ztachip_arch}(a)-(c). The parser extracts operator options like tensor shapes, strides, padding, activation functions, and quantization parameters from the model graph and converts them into an intermediate representation. The compiler then performs quantization preparation and hardware-aware grouping and tiling before generating tensor programs, which are compiled into RISC-V executable code containing the required accelerator register configurations. The hardware consists of a VexRiscv CPU\cite{vexriscv-github}, a dataplane controller, and a multicore acceleration engine. Each acceleration core contains four pcores, with eight ALU lanes per pcore for parallel vector computation. The CPU configures and launches operators, while the dataplane controller and stream processor manage tensor transfers between external DDR memory and pcore-local memories. In this work, we retain the original execution and data-movement flow, extend the baseline per-tensor quantization scheme to per-channel quantization, and add a post-MAC requantization datapath to improve inference accuracy.


\subsection{Per-Channel Quantization Mapping}
The baseline ztachip platform supports only per-tensor quantization, in which all output channels of a convolution layer share a single weight scale. As discussed in \cite{krishnamoorthi2018quantizing}, assigning an independent scale to each output channel better accommodates channel-wise variations in the weight distribution and generally improves inference accuracy. \Cref{eq:float_conv,eq:quant_sub,eq:requant} describe how floating-point convolution is converted to quantized integer computation. The floating-point output of channel $c$ is obtained by accumulating the products of the input activations and the corresponding weights, followed by the addition of the bias. By substituting the quantization relation $F=S(Q-Z)$, the convolution can then be expressed in \Cref{eq:quant_sub} format. Here, $S_a$, $S_{w,c}$, and $S_o$ denote the input-activation scale, the weight scale of output channel $c$, and the output-activation scale, respectively. $Z_a$ and $Z_o$ are the input and output zero-points, while $ACC_c$ represents the accumulated integer convolution result, including the quantized bias. TensorFlow Lite uses symmetric quantization for convolution weights, so the weight zero-point is zero and is omitted from the formulation. The effective scaling factor $S_aS_{w,c}/S_o$ is approximated by a fractional multiplier $M_c$ and a power-of-two shift $2^{-N_c}$, yielding the final requantization form in the entire integer domain. In per-tensor quantization, $S_w$, $M$, and $N$ are shared across all output channels, whereas in per-channel quantization, they are independently assigned to each channel.


\begin{align}
&F_{o,c}
=
\sum_i F_{w,i,c}F_{a,i}+F_{b,c}
\label{eq:float_conv}
\\
&S_o(Q_{o,c}-Z_o)
=
S_aS_{w,c}
\left[
\sum_i(Q_{a,i}-Z_a)Q_{w,i,c}+Q_{b,c}
\right]
\label{eq:quant_sub}
\\
&Q_{o,c}
=
M_c2^{-N_c}\cdot ACC_c+Z_o
\label{eq:requant}
\end{align}


To support this quantization scheme, we redesign the quantization-preparation stage in the ztachip compilation process for operators with multiple weight channels. In the extended flow, the parser first determines whether the weight scale is scalar or channel-wise according to the number of scale values. If it's channel-wise, the compiler generates a corresponding multiplier--shift pair $(M_c,N_c)$ for every output channel. The multiplier and shift vectors are then packed according to the grouping and tiling configuration used by the convolution kernel. Different SIMD ALU lanes will now be configured to apply different multiplier and shift values to their post-MAC results. To enable end-to-end evaluation on a broader range of CNN models, we additionally introduce support for several previously unsupported operations to improve model compatibility, including padding, max pooling, and mean reduction, and extend existing convolution kernels to cover additional kernel sizes and stride configurations.
\subsection{Post-MAC Quantization Datapath}
Executing the TensorFlow Lite-equivalent requantization procedure on the RISC-V CPU would preserve the reference arithmetic. However, it would require frequent transfers of intermediate accumulator values between the PCORE local memory and external memory, resulting in substantial data-movement overhead. ZTA-Q therefore performs post-MAC requantization inside the PCORE acceleration engine. The channel-wise multiplier and shift vectors generated by the compiler are loaded into the pcore together with each output-channel tile. After the SIMD ALU lanes complete the multiply--accumulate operations, the 32-bit accumulation result $ACC_c$, stored in the accumulator register \textit{xreg}, is rescaled using the corresponding multiplier $M_c$ and shift $N_c$. The resulting value is subsequently passed to the stream processor, where the output zero-point $Z_o$, activation function, and any required output remapping are applied.

\begin{equation}
\begin{alignedat}{2}
&Q_{m,c}
=
\left(
ACC_c \times M_{i16,c}
+
round_{m,c}
\right)
\gg 15,
\\
&Q_{s,c}
=
\left(
Q_{m,c}
+
round_{s,c}
\right)
\gg N_c .
\end{alignedat}
\label{rounding computation}
\end{equation}

To retain sufficient fractional precision, the real-valued multiplier $M_c$ is scaled by $2^{15}$ and stored as a signed 16-bit integer $M_{i16,c}$. The quantization procedure is implemented in two rounding stages shown in \Cref{rounding computation}. The first stage removes the $2^{15}$ fixed-point scaling introduced when representing $M_c$ as $M_{i16,c}$. The second stage applies the channel-specific quantization shift. Both are executed in the vector ALU datapath shown in \Cref{ztachip_arch}(d), using the accumulator from \textit{xreg} together with per-channel multipliers and shifts for each SIMD lane. For multiplier rounding, as the product needs to shift right by 15 bits, the offset $round_{m,c}$ is set to $2^{14}$ for a non-negative product and $1-2^{14}$ for a negative product. The second stage derives its rounding offset from the sign, rounding, and sticky bits, where the sticky bit is the OR reduction of the discarded lower bits. As shown in \Cref{subsec::ablation}, preserving the rounding stages, particularly shift rounding, is essential for preventing small numerical deviations from accumulating across the network. 



\section{Experimental Results}
In this section, we first analyze the impact of optional approximation approaches on model accuracy, inference time, and hardware cost. Then, we evaluate the model inference performance on the FPGA board with quantized TensorFlow Lite models.

\subsection{Ablation study of approximation approaches}\label{subsec::ablation}

\begin{figure}[h!]
    \centering
    \includegraphics[width=0.48\textwidth]{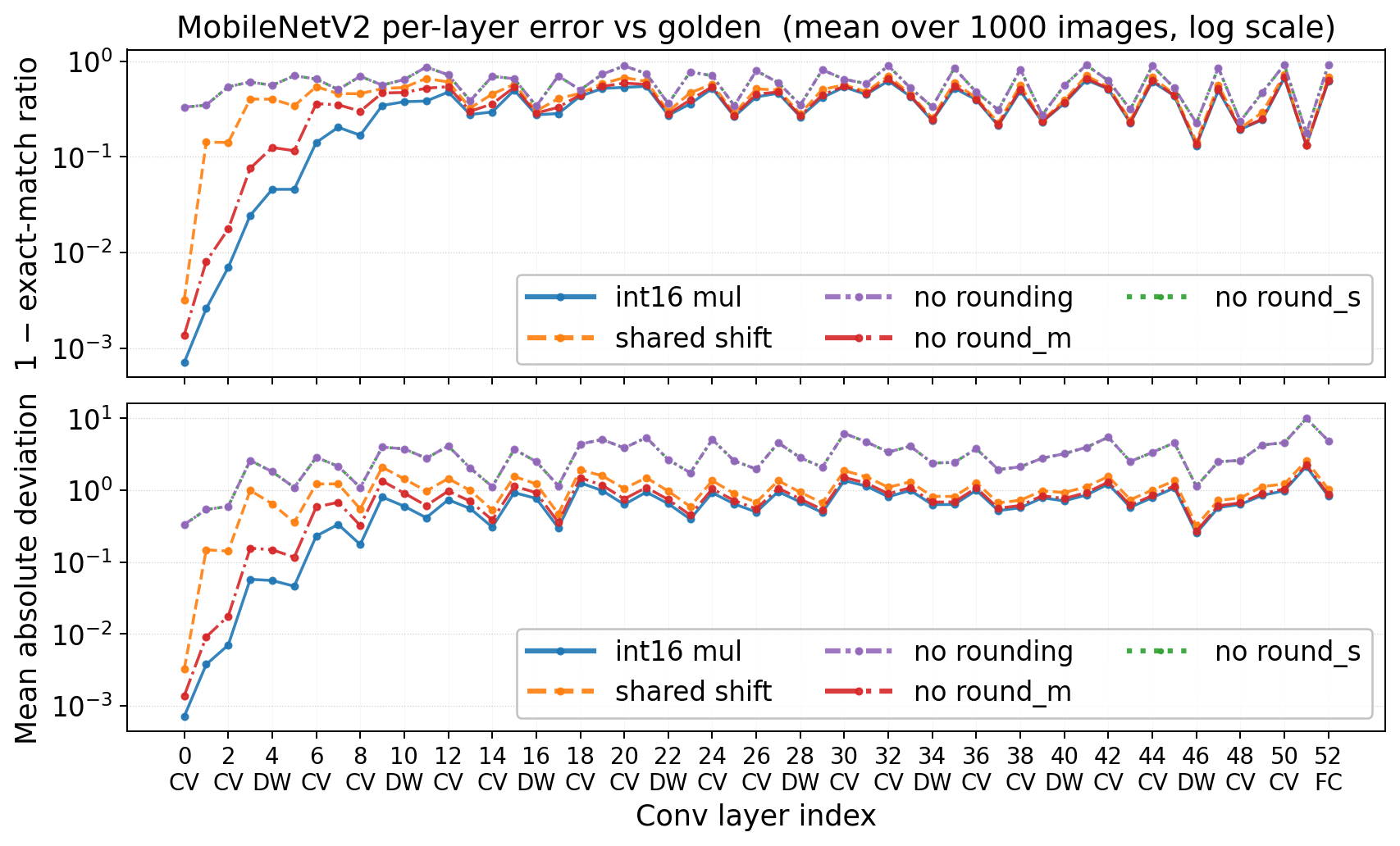}
    \caption{Impact of individual hardware approximations on all types of computational layers in MobileNetV2 over 1000 ImageNet validation images: (a) 1 - exact-match ratio and (b) mean absolute deviation relative to TensorFlow Lite.}
    \label{ablation study}
\end{figure}

Although quantized inference is widely supported, prior accelerator studies rarely examine how circuit-level approximations in the post-MAC quantization datapath affect end-to-end accuracy and hardware cost. Prior works typically report numerical precision and provide a high-level representation of the quantization module in the hardware architecture, without detailing their implementation. We therefore perform an ablation study on all MobileNetV2 convolution layers using five approximations: reduced multiplier ($M_c$) precision, shared shifts with rescaled multipliers, removal of multiplier rounding($round\_m$), removal of shift rounding($round\_s$), and removal of both rounding stages. Here, shared shifts means $N_c$ in \Cref{rounding computation} is shared among channels. As \Cref{ablation study} shows, over a balanced subset of 1,000 ImageNet validation images, with one image randomly selected from each class, the exact-match ratios of different schemes become increasingly similar after successive layers, while their numerical deviations remain distinct. Removing rounding produces substantially larger mean absolute deviation from the TensorFlow Lite reference across convolution (CV), depthwise convolution (DW), and fully connected (FC) layers, whereas INT16 multipliers and shared shifts introduce only minor errors. Notably, the no-shift-rounding curve nearly overlaps with the no-rounding curve, whereas removing only multiplier rounding causes a much smaller deviation. This confirms that shift rounding is the dominant factor in preserving numerical fidelity during post-MAC quantization.


\vspace{-3mm}

\vspace{-5mm}

\begin{table}[h!]
    \centering
    \caption{MobileNetV2 Top-1/Top-5 accuracy with different approximation approaches on the 1000-image subset}
    \label{approx_model_test}
    \begin{tabular}{l|c c c c}
        \hline
        \textbf{Version} & \textbf{Top-1} & \textbf{Top-5}  & \textbf{Inference Time[ms]} \\
        \hline
        TF Float & 71.8\%  & 91.5\% & - \\ 
        TF Lite & 72.0\% & 91.6\%  & - \\ 
        V0 & 67.3\% (-4.7 pp) & 89.2\% (-2.4 pp) & 178.7 \\ 
        V1 & 67.0\% (-5.0 pp) & 89.3\% (-2.3 pp) & 170.0 \\ 
        \textit{\textbf{ZTA-Q}} & 71.0\% (-1.0 pp) & 91.1\% (-0.5 pp) & 170.0 \\ 
        \hline
    \end{tabular}
\end{table}

\vspace{-6mm}

\begin{table}[h!]
    \centering
    \caption{FPGA Resource Utilization with different approximation approaches}
    \label{utilization}
    \begin{tabular}{l|c c c c | c}
        \hline
        \textbf{Resource} & Baseline & V0  & V1 & \textbf{\textit{ZTA-Q}} & \textbf{Overhead} \\
        \hline
        Slice LUT       & 37320 & 39706 & 42510 & 47126 & 26.3\% \\ 
        Slice Register  & 46251 & 50458 & 50728 & 52084 & 12.6\% \\ 
        F7 MUX          & 1677  & 2477  & 1545  & 1549  & -7.6\% \\ 
        F8 MUX          & 505   & 505   & 505   & 505   & 0 \\ 
        BRAM            & 85.5  & 85.5  & 85.5  & 85.5  & 0 \\ 
        DSP             & 64    & 160   & 160   & 160   & 150\% \\ 
        \hline
    \end{tabular}
\end{table}

\begin{table*}[t]
    \centering
    \caption{Evaluation of common CNN models on ImageNet and MNIST}
    \label{cnn model test}
    \begin{tabular}{l|ccc ccc c}
        \toprule
        \multirow{2}{*}{\textbf{Model}} 
        & \multicolumn{3}{c}{\textbf{Top-1}} 
        & \multicolumn{3}{c}{\textbf{Top-5}} 
        & \multirow{2}{*}{
    \makecell{\textbf{Inference Time per Image}\\
              \textbf{(Arty A7-100T@83.3MHz)}}} \\
        \cmidrule(lr){2-4} \cmidrule(lr){5-7} 
        & \textbf{TF Float} & \textbf{TF Lite} & \textbf{ZTA-Q}
        & \textbf{TF Float} & \textbf{TF Lite} & \textbf{ZTA-Q}
        & \\
        \midrule
        LeNet-5 
        & 99.20\% & 99.19\% & 99.19\% (0 pp)
        & 100\% & 100\% & 100\% (0 pp)
        & 18.2 ms\\

        SqueezeNet1.1 
        & 58.19\% & 57.95\% & 57.72\% (-0.23 pp)
        & 80.62\% & 80.40\% & 80.22\% (-0.18 pp)
        & 331.2 ms \\

        MobileNetV1 
        & 70.32\% & 68.95\% & 68.95\% (0 pp)
        & 89.44\% & 88.56\% & 88.57\% (+0.01 pp)
        & 256.3 ms \\
        
        MobileNetV2 [UINT8] 
        & -- & 71.12\% & 71.10\% (-0.02 pp)
        & -- & 89.96\% & 89.88\% (-0.08 pp)
        & 170.0 ms \\
        
        MobileNetV2 [INT8] 
        & 71.48\% & 70.72\% & 70.68\% (-0.04 pp)
        & 90.27\% & 89.88\% & 89.90\% (+0.02 pp)
        & 170.0 ms \\
        
        ResNet-18 
        & 69.81\% & 69.57\% & 69.50\% (-0.07 pp)
        & 89.11\% & 88.93\% & 88.91\% (-0.02 pp)
        & 1.0 s \\

        ResNet-50
        & 74.72\% & 74.58\% & 74.36\% (-0.22 pp)
        & 91.96\% & 91.79\% & 91.69\% (-0.10 pp)
        &  2.4 s \\
        
        \bottomrule
    \end{tabular}
\end{table*}

We then consolidate the five approximations into three hardware versions: V0 uses per-channel multipliers with a shared shift, V1 introduces per-channel shifts, and ZTA-Q further adds two-stage rounding. As shown in \Cref{approx_model_test}, V0 reduces top-1 and top-5 accuracy by 4.7 and 2.4 percentage points, while per-channel shifts in V1 provide only marginal improvement. In contrast, ZTA-Q restores accuracy close to the TensorFlow Lite reference, confirming that rounding is the dominant accuracy factor among the evaluated hardware approximations. All designs are implemented on the Arty A7-100T at 83.3 MHz using the \textit{\textbf{Performance\_NetDelay\_High}} strategy. As reported in \Cref{utilization}, ZTA-Q increases LUT, register, and DSP utilization by 26.3\%, 12.6\%, and 150\%, respectively, while enabling TensorFlow Lite quantized models with negligible accuracy degradation. Here, the substantial increase in DSP utilization is mainly caused by replacing the original int16 multiplier with an int32 multiplier to support the multiplication between the accumulator value and the quantization multiplier. 

\subsection{End-to-End CNN model inference evaluation}
Following the ablation study, we evaluate our implementation across representative popular CNN models. To cover representative edge-inference workloads with different network depths and operator compositions, we select LeNet-5, SqueezeNet1.1, MobileNetV1/V2, ResNet-18 and ResNet-50 as benchmarks, using full MNIST and ImageNet as their corresponding validation datasets. For post-training quantization, the representative dataset contains 1,000 calibration images, with an equal number of images randomly selected from each class, and the test dataset is the full ImageNet and MNIST validation set. The provided MobileNetV2 model in the ztachip GitHub repository uses unsigned 8-bit activations and per-tensor quantization, whereas the remaining models are quantized using signed INT8 activations and per-channel quantization. The evaluated models and batch evaluation scripts are made publicly available on GitHub\footnote{https://github.com/kevinli-hw/ZTA-Q}. As shown in \Cref{cnn model test}, the proposed implementation achieves an accuracy difference of less than 0.25 percentage points compared with the corresponding TensorFlow Lite inference results across all evaluated models.

\section{Conclusion}

In this work, we presented ZTA-Q, an open-source RISC-V-based platform for accurate quantized CNN inference on edge FPGAs. Guided by an ablation study of hardware approximations, we develop a post-MAC quantization flow with two-stage rounding for both per-tensor and per-channel schemes, achieving negligible accuracy degradation on TensorFlow Lite models. Future work will focus on evaluating broader workloads and quantization schemes, including attention, mixed precision, dynamic quantization, and non-uniform per-channel parameters.

\bibliographystyle{IEEEtran}
\bibliography{reference}

\end{document}